\documentstyle[12pt]{article}
\begin{document}
\begin{flushright}
hep-th/0402005\\
SNB/February/2004
\end{flushright}
\vskip 2cm
\begin{center}
{\bf \Large { Nilpotent Symmetries For Matter Fields \\
In Non-Abelian Gauge Theory: Augmented Superfield Formalism}}

\vskip 3cm

{\bf R.P.Malik}\\
{\it S. N. Bose National Centre for Basic Sciences,} \\
{\it Block-JD, Sector-III, Salt Lake, Calcutta-700 098, India} \\
{\bf E-mail address: malik@boson.bose.res.in  }\\

\vskip 2.5cm

\end{center}

\noindent
{\bf Abstract}:
In the framework of superfield approach to Becchi-Rouet-Stora-Tyutin (BRST)
formalism, the derivation of the 
(anti-)BRST nilpotent symmetry transformations for the matter
fields, present in any arbitrary  interacting gauge theory,
has been a long-standing problem. 
In our present investigation, the local, covariant, continuous and off-shell 
nilpotent (anti-)BRST symmetry transformations
 for the Dirac fields $(\psi, \bar\psi)$ are derived in the 
framework of the augmented
superfield formulation where the four ($3 + 1$)-dimensional (4D)
interacting non-Abelian gauge 
theory is considered on the six $(4 + 2)$-dimensional supermanifold 
parametrized by the four {\it even} spacetime coordinates $x^\mu$ and a couple 
of {\it odd} elements ($\theta$ and $\bar\theta$) of the Grassmann algebra. The
requirement of the invariance of the matter (super)currents and the 
horizontality
condition on the (super)manifolds leads to the derivation of the nilpotent 
symmetries for the matter fields as well as the gauge- and the (anti-)ghost
fields of the theory in the general scheme of augmented
superfield formalism.\\

\vskip 0.5cm



\baselineskip=16pt

%\vskip 1cm

\newpage

\noindent
{\bf 1 Introduction}\\

\noindent
The symmetry groups and corresponding algebras have played some notable roles
in the developments of the modern theoretical high energy physics up to the
energy scale of the order of Planck's scale. In particular,
the local and continuous symmetry groups corresponding to the gauge theories
\footnote{This set of theories is endowed with the first-class constraints 
in the language of the Dirac's prescription for the classification of 
constraints [1,2]. This observation
is true for any arbitrary $p$-form gauge theories.} 
have been found to dictate three (out of four) fundamental interactions of 
nature which govern physics up to the energy scale
of the order of grand unification. The
requirement of the {\it local} gauge invariance 
(i.e. gauge principle) enforces the existence of an interaction
term in the Lagrangian density of the theory where the gauge field couples to
the matter conserved current of the interacting gauge theory [3]. 
The existence of the conserved matter current owes its origin to
the presence of a global gauge invariance in the theory
(Noether's theorem). One of the
most elegant and intuitive methods of covariantly quantizing such a class of
gauge theories is the Becchi-Rouet-Stora-Tyutin (BRST) formalism where
the unitarity and the ``quantum'' gauge (i.e. BRST) invariance are respected 
together at any arbitrary order of perturbative computations for a given
physical process. The reach and range of the usefulness of this 
intuitive formalism have been beautifully extended to include the second-class 
constraints in its domain of applicability [4]. This formalism has now been 
found to hold a firm ground in the realm of the modern developments in the 
context of topological field theories [5-7],
topological string theories [8], (super)string theories and their close
cousins D-branes and M-theory (see, e.g., [9,10] and references therein).
Its deep connection with the mathematics 
(see, e.g., [11-15] for details) of cohomology and differential
geometry, its very striking similarity with some of the key concepts
of supersymmetry,
its natural inclusion in the Batalin-Vilkovisky scheme [16,17], its intuitive
interpretation in the language of geometry on the supermanifold, etc., have
been responsible for this active topic of research to soar a fairly high 
degree of mathematical sophistication and very useful (as well as
attractive) physical applications.

In our present investigation, we shall be concentrating on the geometrical
aspects of the BRST formalism in the framework of the augmented superfield
formalism. Such a study is expected to shed some light on a few abstract
mathematical structures behind the BRST formalism in a more intuitive
and transparent manner.
The usual superfield approach [18-25] to BRST formalism
delves deep into providing
the geometrical origin and interpretation for the conserved
and nilpotent ($Q_{(a)b}^2 = 0$) (anti-)BRST charges 
($Q_{(a)b}$) which generate a set of local, covariant, continuous, nilpotent 
$(s_{(a)b}^2 = 0$) and anticommuting $(s_b s_{ab} + s_{ab} s_b = 0$)
(anti-)BRST symmetry transformations $(s_{(a)b})$
\footnote{We shall be following the notations and conventions adopted by
Weinberg [26]. In its full glory, a nilpotent ($\delta_{B}^2 = 0$) BRST
symmetry transformation $\delta_B$ is the product of an anticommuting
spacetime independent parameter $\eta$ and $s_b$
as $\delta_B = \eta s_b$ with $s_b^2 = 0$. The parameter $\eta$ anticommutes
with all the fermionic fields (e.g. Dirac fields, (anti-)ghost fields, etc.)
of a given interacting $p$-form gauge theory.} {\it only} for
the gauge field and the (anti-)ghost fields of the Lagrangian
density of a given $p$-form $ (p = 1, 2, 3....)$ {\it interacting} 
Abelian gauge theory
in the $D$-dimensions of spacetime. In this formalism [18-25], 
one constructs a $(p + 1)$-form super curvature $\tilde F = \tilde d \tilde A$ 
by exploiting
the super exterior derivative $\tilde d$
(with $\tilde d^2 = 0$) and the super connection $p$-from $\tilde A$
on a $(D + 2)$-dimensional supermanifold parametrized by $D$-number of
spacetime even coordinates $x^\mu  (\mu = 0, 1, 2...D-1)$
and a couple of odd coordinates $\theta, \bar \theta$
(with $\theta^2 = \bar\theta^2 = 0, \theta\bar\theta + \bar\theta \theta = 0$)
of the Grassmann algebra. This super curvature $(p + 1)$-form $\tilde F$
is subsequently equated, due to the so-called horizontality condition
\footnote{This is referred to as the soul-flatness condition in [27] which
amounts to setting equal to zero all the Grassmannian components of the
$(p + 1)$-rank (anti-)symmetric super curvature tensor defined on the 
$(D + 2)$-dimensional supermanifold corresponding to a given
$p$-form Abelian gauge theory.}, with the $(p + 1)$-form ordinary curvature
$F = d A $ constructed from the ordinary exterior derivative
$d = dx^\mu \partial_\mu$ (with $d^2 = 0$) and ordinary $p$-form
connection $A$ of a given $p$-form Abelian gauge theory
\footnote{For the 1-form 
($A = dx^\mu A_\mu \cdot T$) non-Abelian gauge theory (see, e.g. [20]), 
the super
2-form curvature $\tilde F = \tilde d \tilde A + \tilde A \wedge \tilde A$
(defined on the six $(4 + 2)$-dimensional supermanifold)
is equated with the ordinary 2-form curvature $F = d A + A \wedge A$
(defined on the ordinary four $(3 + 1)$-dimensional Minkowskian submanifold
of the above supermanifold)
in the horizontality condition $\tilde F = F$ which leads to the
exact derivation of the nilpotent and anticommuting (anti-)BRST
symmetry transformations associated
with the gauge (e.g. $s_b A_\mu = D_\mu C, s_{ab} A_\mu = D_\mu \bar C$) and 
(anti-)ghost fields (e.g. $s_b C = \frac{1}{2} C \times C, s_b \bar C = i B,
s_{ab} C = i \bar B, s_{ab} \bar C = \frac{1}{2} \bar C \times \bar C$)
of the four $(3 + 1)$-dimensional non-Abelian gauge theory 
(see, e.g., Sec. 3 below for the details).}. 
In fact, the process of
reduction of the (anti-)symmetric second-rank super tensor $\tilde F_{MN}$
(of $\tilde F$) to the ordinary 
antisymmetric second-rank tensor $F_{\mu\nu}$ (of $F$) 
produces the nilpotent
(anti-)BRST symmetry transformations for the gauge- and the (anti-)ghost
fields of the interacting $p$-form Abelian gauge theory. This restriction
(i.e. $\tilde F = F$), however, does not shed any light on the nilpotent 
and anticommuting (anti-)BRST symmetry transformations for the {\it matter}
fields of the interacting $p$-form Abelian gauge theory. Thus, it is 
clear that
the derivation of the nilpotent symmetry transformations for {\it all}
 the fields, present in a given general $p$-form {\it interacting} 
(non-)Abelian gauge theory, is incomplete as far as the above superfield 
formulation [18-25] is concerned.

In our recent set of 
papers [28-31], the above usual superfield formulation [18-25,27] has been
augmented by invoking the invariance of the conserved matter (super)currents
on the (super)manifolds which leads to the derivation of the
off-shell nilpotent symmetry transformations for the matter fields of
the interacting $1$-form gauge theory. We christen this extended version of
the superfield formulation [28-31] as the augmented superfield formalism.
The most interesting part
of this prescription is the fact that there is a mutual consistency
and complementarity between the requirements of (i) the horizontality 
condition, and (ii) the invariance of the matter (super)currents.
The latter requirement is not imposed by hand 
from outside as is the case with the former
(i.e. the horizontality condition). Rather, it turns 
out to be the inherent property of the interacting gauge theory itself.
As the prototype examples of such a class of interacting
gauge theories, we have chosen the 1-form interacting Abelian theories
where (i) the Dirac fields couple to the Abelian gauge field $A_\mu$ in
two $(1 + 1)$-dimensional (2D) spacetime, and (ii) the
complex scalar fields and the Dirac fields couple to the Abelian $U(1)$ gauge 
field $A_\mu$ in four $(3 + 1)$-dimensions (4D) of spacetime. In the above
field theoretical examples (i) and (ii), the equality (i.e.
$\tilde J_\mu (x,\theta,\bar\theta) = J_\mu (x)$) of the matter 2- and 4-vector
supercurrent $\tilde J_\mu (x,\theta,\bar\theta)$ 
(defined on the $(2 + 2)$- and $(4 + 2)$-dimensional supermanifolds, 
respectively) with the ordinary matter
2- and 4-vector conserved current $J_\mu (x)$ 
(defined on the $(1 + 1)$- and $(3 + 1)$-dimensional 
ordinary spacetime manifolds) leads to the derivation of
the nilpotent (anti-)BRST
symmetry transformations for the matter (i.e. the Dirac and the complex
scalar) fields of the above chosen interacting gauge theories. These are found
to be consistent with the nilpotent (anti-)BRST transformations for
the gauge- and the (anti-)ghost fields derived from the horizontality
condition. Furthermore, for the case of the field theoretic example in
(i), the invariance of the axial-vector (super)currents
(i.e. $\tilde J_\mu^{(5)} (x,\theta,\bar\theta) = J_\mu^{(5)} (x)$)
on the $(2 + 2)$-dimensional (super)manifolds leads to the derivation
of the nilpotent (anti-)co-BRST symmetry transformations for the Dirac fields
which are checked to be consistent with the anticommuting
and nilpotent (anti-)co-BRST transformations
derived for the gauge- and the (anti-)ghost fields from the dual-horizontality
condition (see, e.g., [28] for details).

The purpose of the present paper is to derive, in the framework
of the augmented superfield formulation,  the off-shell nilpotent
version of the (anti-)BRST symmetry transformations for the Dirac (matter)
fields of an  interacting 
non-Abelian gauge theory where there is a coupling between
the conserved matter vector current $J_\mu (x) = \bar \psi \gamma_\mu \psi$
and the group valued (i.e. $ A_\mu = A_\mu^a T^a \equiv A_\mu \cdot T$) 
non-Abelian gauge field $A_\mu$. We derive the off-shell (as well as on-shell)
nilpotent symmetry transformations for the gauge- and the (anti-)ghost fields 
by exploiting the horizontality condition
and the corresponding nilpotent symmetries for the matter fields are
derived by invoking the invariance of the matter (super)currents defined
on the (super)manifolds. We also derive separately (and independently)
the BRST and anti-BRST symmetry transformations for the matter fields by
invoking the invariance of the matter conserved (super)currents on the
$(4 + 1)$-dimensional (anti-)chiral super sub-manifolds embedded
in the most general $(4 + 2)$-dimensional supermanifold.
The geometrical interpretation for the 
nilpotent (anti-)BRST charges $Q_{(a)b}$ as the translation generators
on the six $(4 + 2)$-dimensional supermanifold is exactly same as the
case of interacting Abelian gauge theories (see, e.g., [28,29]).
The above interpretation is valid for the translations of all the superfields
$(B_\mu, \Phi, \bar\Phi, \Psi, \bar \Psi)$ 
(cf.(3.1) and (4.1) below) corresponding to the gauge field,
the (anti-)ghost fields and the matter fields  (i.e. $A_\mu, C, \bar C, \psi, 
\bar\psi$) of the Lagrangian density (cf.(2.4) below)
for the given 1-form interacting non-Abelian gauge theory in 4D.
Our present study is essential primarily on three
counts. First, it has been an outstanding problem to derive the
nilpotent symmetries for the matter fields in the framework of
superfield formulation. Second, through this work, we generalize our
earlier works [28,29] on interacting Abelian gauge theories to the more
general case of interacting non-Abelian gauge theory as the former is
a limiting case of the latter. Third, to check the mutual consistency
between the horizontality condition and the
invariance of the matter (super)currents for
the case of non-Abelian gauge theory which was found to be true for
the case of interacting Abelian gauge theories in 2D and 4D (see, e.g.,
[28,29]).

The contents of our present paper are organized as follows. 
To set up the notations and conventions, in section 2,
we briefly sketch the off-shell (as well as the on-shell)
nilpotent (anti-)BRST symmetry transformations for all
the fields present in the Lagrangian density of an interacting 
non-Abelian gauge theory in four $(3 + 1)$-dimensions of spacetime.
For the sake of the present paper to be self-contained, in section 3,
we derive the off-shell (as well as the on-shell)
nilpotent symmetry transformations for the gauge- 
and the (anti-)ghost fields  in the framework of 
the usual superfield formulation by
exploiting the horizontality condition (see, e.g.,[20,32,33] for details). 
The central result of the derivation of the nilpotent
(anti-)BRST symmetry transformations for the Dirac fields
is contained in section 4 where we exploit the invariance of the matter
(super)currents on the most general (super)manifolds
and the (anti-)chiral super sub-manifolds. Finally, we make some 
concluding remarks in section 5 and point out a few future directions 
which could be pursued later.\\

\noindent
{\bf 2 Nilpotent (Anti-)BRST Symmetries: Lagrangian Formalism}\\

\noindent
For the sake of simplicity,
let us begin with the BRST invariant Lagrangian density ${\cal L}_{b}$
for the interacting 4D non-Abelian gauge theory 
\footnote{We follow here the conventions and notations such that the flat
4D Minkowski metric
$ \eta_{\mu\nu} = $ diag $ (+1, -1, -1, -1), 
D_{\mu} C = \partial_{\mu} C + A_{\mu} \times C, 
\alpha \cdot \beta = \alpha^{a} \beta^{a},
(\alpha \times \beta)^a = f^{abc} \alpha^{b} \beta^{c}$ where $\alpha$
and $\beta$ are the non-null vectors in the group space. 
Here the Greek indices: $\mu, \nu, \rho...= 0, 1, 2, 3$ correspond to the
spacetime directions and Latin indices: 
$ a, b, c...= 1, 2, 3...$ stand for the ``colour'' values in the group space.} 
in the Feynman gauge [26,27,34-36]
$$
\begin{array}{lcl}
{\cal L}_{b} = - \frac{1}{4}\; F^{\mu\nu}\cdot F_{\mu\nu} 
+ \bar \psi \;(i \gamma^\mu D_\mu - m) \; \psi 
+ B \cdot (\partial_{\rho} A^{\rho})
+ \frac{1}{2}\; B \cdot B
- i \;\partial_{\mu} \bar C \cdot D^\mu C, 
\end{array} \eqno(2.1)
$$
where $F_{\mu\nu} = \partial_{\mu} A_{\nu} -
\partial_{\nu} A_{\mu} + (A_{\mu} \times A_{\nu})$ is the field 
strength tensor derived from the one-form connection $ A = d x^\mu A_\mu \equiv
dx^\mu A^{a}_{\mu} T^{a} = dx^\mu (A_\mu \cdot T)$ 
by the Maurer-Cartan equation $F = d A + A \wedge A$ 
with $ F = \frac{1}{2} d x^\mu \wedge d x^\nu F_{\mu\nu}^{a} T^{a}$. Here
$T^{a}$ are the generators of the compact Lie algebra $ [ T^{a}, T^{b} ]
= i f^{abc} T^{c}$ where $f^{abc}$ are the structure constants that can be
chosen to be totally antisymmetric in $ a, b, c$ 
(see, e.g., [26] for details). The anticommuting
($ (C^a)^ 2 = (\bar C^a)^2 = 0, C^a \bar C^b + \bar C^b C^a = 0,
C^a \psi + \psi C^a = 0$ etc.) (anti-)ghost
fields $(\bar C^a) C^a$ (which interact with 
the self-interacting non-Abelian gauge fields $A_{\mu}$ only
in the loop diagrams) are required to be present in the theory to maintain the
unitarity and the ``quantum'' gauge 
(i.e. BRST) invariance together at any arbitrary order of perturbative
computations [37]. These fields (even though 
they do interact with the gauge fields $A_\mu$) are not the physical 
matter fields. The physical matter fields of the theory are the Dirac (quark)
fields $(\bar\psi, \psi)$ with the covariant derivative defined as
\footnote{ Note that the coupling constant $g$ in the actual definition of
the covariant derivative $D_\mu \psi = \partial_\mu \psi + i g A_\mu \psi,
D_\mu C = \partial_\mu C + g A_\mu \times C$ has been set equal to
one (i.e. $ g = 1$) for the sake of brevity.}
: $D_\mu \psi = \partial_\mu \psi
+ i A_\mu \psi$ where the gauge field $A_\mu$ is group valued
(i.e $A_\mu = A_\mu \cdot T$).
The above Lagrangian density (2.1)
respects the following off-shell 
nilpotent ($s_{b}^2 =0 $) BRST ($s_{b}$)
symmetry transformations [26,27,34-36]
$$
\begin{array}{lcl}
&& s_{b} A_\mu = D_\mu C, \qquad s_{b} C = + \frac{1}{2} C \times C, \qquad
s_{b} \bar C = i B, \qquad s_b B  = 0, \nonumber\\ 
&& s_{b} \psi = -i (C \cdot T) \psi, \qquad s_b \bar \psi = - i
\bar\psi (C \cdot T), \qquad s_b F_{\mu\nu} = F_{\mu\nu} \times C.
\end{array} \eqno(2.2)
$$
It is worth pointing out, at this juncture, that 
(i) the kinetic energy term $-\frac{1}{4} 
(F^{\mu\nu} \cdot F_{\mu\nu}$)
remains invariant under the BRST transformation $s_{b}$ because of the 
presence of a totally antisymmetric $f^{abc}$ in its variation (i.e.
$s_b (-\frac{1}{4} F^{\mu\nu} \cdot F_{\mu\nu}) = - \frac{1}{2}
f^{abc} F^{a\mu\nu} F_{\mu\nu}^b C^c = 0$). 
(ii) There exists an on-shell ($\partial_\mu D^\mu C = 0$) nilpotent
$(\tilde s_b^2 = 0$) version of the above BRST symmetry transformations
for this interacting non-Abelian gauge theory (see, e.g., [33])
$$
\begin{array}{lcl}
&& \tilde s_{b} A_\mu = D_\mu C, \qquad\;\; 
\tilde s_{b} C = + \frac{1}{2} C \times C, \qquad\;\;
\tilde s_{b} \bar C = - i (\partial_\mu A^\mu), \nonumber\\ 
&& \tilde s_{b} \psi = -i (C \cdot T) \psi, \qquad 
\tilde s_b \bar \psi = - i
\bar\psi (C \cdot T), \qquad \tilde s_b F_{\mu\nu} = F_{\mu\nu} \times C,
\end{array} \eqno(2.3)
$$
under which, the following Lagrangian density for the interacting non-Abelian
gauge theory
$$
\begin{array}{lcl}
{\cal L}_{\tilde b} = - \frac{1}{4}\; F^{\mu\nu}\cdot F_{\mu\nu} 
+ \bar \psi \;(i \gamma^\mu D_\mu - m) \; \psi 
-\frac{1}{2}\; 
(\partial_\mu A^\mu) \cdot (\partial_{\rho} A^{\rho})
- i \;\partial_{\mu} \bar C \cdot D^\mu C, 
\end{array} \eqno(2.4)
$$
transforms to a total derivative. It will be noted that (2.4) and (2.3) have
been derived from (2.1) and (2.2), respectively,
by exploiting the equation of motion
$B + (\partial_\mu A^\mu) = 0$ emerging from (2.1).
(iii) Besides the symmetry transformations (2.2) and (2.3), there
exists a nilpotent ($s_{ab}^2 = 0$) anti-BRST symmetry 
transformation ($s_{ab}$) that is also present in the theory,
in a subtle way. To realize this anti-BRST symmetry, one has to introduce 
another auxiliary field $\bar B$ 
(satisfying $ B + \bar B = i\; C \times \bar C$) to recast the
Lagrangian density (2.1) into the following 
equivalent forms (see, e.g., [38,34-36])
$$
\begin{array}{lcl}
{\cal L}_{\bar B} 
&=&  - \frac{1}{4}\; F^{\mu\nu} \cdot F_{\mu\nu}
+ \bar \psi (i \gamma^\mu D_\mu - m) \psi
+ B \cdot (\partial_{\mu}  A^{\mu}) 
\nonumber\\
&+& \frac{1}{2} (B \cdot B + \bar B \cdot
\bar B) - i \partial_{\mu} \bar C \cdot D^\mu C,
\end{array} \eqno(2.5a)
$$
$$
\begin{array}{lcl}
{\cal L}_{\bar B}
&=&  - \frac{1}{4}\; F^{\mu\nu} \cdot F_{\mu\nu} + \bar \psi
(i \gamma^\mu D_\mu - m) \psi
- \bar B \cdot (\partial_{\mu}  A^{\mu}) \nonumber\\
&+& \frac{1}{2} (B \cdot B + \bar B 
\cdot \bar B) - i D_{\mu} \bar C \cdot \partial^\mu C.
\end{array} \eqno(2.5b)
$$
The Lagrangian density (2.5b) transforms to a total derivative under the 
following off-shell nilpotent ($s_{ab}^2 = 0$)
anti-BRST ($s_{ab}$) symmetry
transformations (see, e.g., [34,26] for details)
$$
\begin{array}{lcl}
 s_{ab} A_{\mu} &=& D_{\mu} \bar C, \qquad \;s_{ab} \bar C 
= + \frac{1}{2} \bar C \times \bar C, \qquad\; 
s_{ab}  C = i \bar B, \qquad  \;s_{ab} B =  B \times \bar C, \nonumber\\
s_{ab} F_{\mu\nu}  &=& F_{\mu\nu} \times \bar C, \quad s_{ab} \psi =
- i (\bar C \cdot T) \psi, \quad \;s_{ab} \bar\psi = - i \bar \psi
(\bar C \cdot T), \quad \;s_{ab} \bar B =  0. 
\end{array} \eqno(2.6)
$$
It can be explicitly verified, using (2.2) and (2.6), that
$s_b s_{ab} + s_{ab} s_b = 0$ for any arbitrary field of (2.1).
For such a verification, the transformation (2.2) 
should be extended to include $s_b \bar B = \bar B \times C$. At this stage,
it is worthwhile to point out that, unlike the BRST symmetries
(cf. (2.2) and (2.3)), there is
no on-shell nilpotent version of the anti-BRST symmetry transformations
in (2.6). All the above continuous symmetry
transformations can be concisely expressed, in terms of the Noether
conserved and the off-shell (as well as on-shell)
nilpotent ($Q_r^2 = 0, \tilde Q_r^2 = 0$) charges $Q_{r} (\tilde Q_{r})$ as 
$$
\begin{array}{lcl}
s_{r} \Sigma (x) = - i \; [ \Sigma (x), Q_{r} ]_{\pm}, \qquad
r = b, ab, \qquad
\tilde s_{r} \Sigma (x) = - i \; [ \Sigma (x), \tilde Q_{r} ]_{\pm}, \qquad
r = b, 
\end{array} \eqno(2.7)
$$
where the $\pm$ signs on the brackets stand for the brackets to be an
(anti-)commutator for the generic field $\Sigma
= A_\mu, C, \bar C, \psi, \bar\psi, B, \bar B$ being (fermionic)bosonic 
in nature. The transformations $\tilde s_r$, in the above,
correspond to the on-shell
nilpotent BRST transformations (2.3).\\

\noindent
{\bf 3 Nilpotent Symmetries for Gauge- and (Anti-)ghost 
Fields: Superfield Formalism}\\

\noindent
We recapitulate here the bare essentials of some of the key points
connected with the idea of the horizontality condition and
its application. To this end in mind,
first of all, let us generalize the generic local field $\Sigma (x) = 
(A_\mu, C, \bar C) (x)$ corresponding to the gauge- and the (anti-)ghost fields
of the 4D Lagrangian density (2.1) to a
supervector superfield $(\tilde A_{M} \cdot T) (x,\theta,\bar \theta)
= (B_\mu \cdot T, \Phi \cdot T, \bar \Phi \cdot T) 
(x, \theta, \bar \theta)$ defined on the six $(4 + 2)$-dimensional 
supermanifold with the following super expansion (see, e.g., [20,32,33] 
for details)
$$
\begin{array}{lcl}
(B_{\mu} \cdot T) (x,  \theta, \bar \theta) &=& (A_{\mu} \cdot T) (x) 
+\; \bar \theta\; (R_{\mu} \cdot T) (x) +  \theta\;
(\bar R_\mu \cdot T) (x)
+ i\;\theta\;\bar\theta \;(S_\mu \cdot T) (x), \nonumber\\ 
(\Phi \cdot T) (x, \theta,  \bar \theta) &=& (C \cdot T) (x) 
+ i\; \bar \theta \;({\cal B} \cdot T) (x) + i\; \theta\;
(\bar B \cdot T) (x) + i \;\theta\;\bar\theta\; (s \cdot T) (x), \nonumber\\
(\bar \Phi \cdot T) (x,  \theta, \bar \theta) &=& (\bar C \cdot T) (x) 
+ i\; \bar \theta \;(B \cdot T) (x) + i\;\theta\; (\bar {\cal B} \cdot T) (x)
+ i\; \theta\;\bar\theta\; (\bar s \cdot T) (x). 
\end{array} \eqno(3.1)
$$
The noteworthy points, at this stage, are (i)
all the superfields $(\tilde A_{M} \cdot T) 
(x,\theta,\bar\theta)
= (B_\mu \cdot T, \Phi \cdot T, \bar \Phi \cdot T) (x,\theta,\bar\theta)$
as well as local fields (e.g. $A_\mu = A_\mu \cdot T, C = C \cdot T$ etc.)
are group valued;  (ii) the degrees of freedom of the group valued
\footnote{ Hereafter, 
 we shall be shuffling between the explicit group
valued notations (e.g. $A_\mu = A_\mu \cdot T, R_\mu = R_\mu \cdot T,
s = s \cdot T$ etc.) and their shorter versions (e.g. $A_\mu, R_\mu, s$ etc.)
in the remaining part of the paper.}
fermionic (odd) fields $R_\mu, \bar R_\mu, C, \bar C, s, \bar s$
match with that of the bosonic (even) fields $A_\mu, S_\mu, B, \bar B,
{\cal B}, \bar {\cal B}$ so that the theory can be consistent with the
basic requirements of supersymmetry; and (iii) the horizontality restriction
$\tilde F = \tilde D \tilde A = D A = F$ (where $\tilde D \tilde A
= \tilde d \tilde A + \tilde A \wedge \tilde A,
D A = d A + A \wedge A$) leads to the following
relationships (see, e.g., [20,32,33] for details)
$$
\begin{array}{lcl}
R_\mu (x) &=& D_\mu C (x), \quad \bar R_\mu (x) = D_\mu \bar C (x),
\quad B (x) + \bar B (x)  = i\; (C \times \bar C) (x), \nonumber\\
{\cal B} (x) &=& - \frac{i}{2}\; (C \times C) (x), \quad 
\bar {\cal B} (x) = - \frac{i}{2} \;(\bar C \times \bar C) (x), \quad 
\bar s(x) = - (B \times \bar C) (x), \nonumber\\
S_\mu (x) &=& D_\mu B (x) - (D_\mu C \times \bar C) (x),
\qquad s(x) = (\bar B \times C) (x), 
\end{array} \eqno(3.2)
$$
where $S_\mu (x)$ can be equivalently written as:
$S_\mu (x) = - D_\mu \bar B (x) - (D_\mu \bar C \times C) (x)$ and 
the individual terms in $\tilde F = \tilde d \tilde A + \tilde A \wedge
\tilde A$ can be explicitly expressed as
$$
\begin{array}{lcl}
\tilde d \tilde A &=& (dx^\mu \wedge dx^\nu) (\partial_\mu B_\nu)
+ (dx^\mu \wedge d\theta) (\partial_\mu \bar\Phi - \partial_\theta B_\mu)
- (d\theta \wedge d \theta) (\partial_\theta \bar\Phi) \nonumber\\
&+& (dx^\mu \wedge d \bar\theta) 
(\partial_\mu \Phi - \partial_{\bar\theta} B_\mu)
- (d\bar\theta \wedge d \bar\theta) (\partial_{\bar\theta} \Phi)
- (d\theta \wedge d \bar\theta) (\partial_\theta \Phi +
\partial_{\bar\theta} \bar \Phi), \nonumber\\
\tilde A \wedge \tilde A &=& (dx^\mu \wedge dx^\nu) (B_\mu B_\nu)
+ (dx^\mu \wedge d\theta) ( [ B_\mu, \bar\Phi ] )
- (d\theta \wedge d \theta) (\bar \Phi \bar\Phi), \nonumber\\
&+& (dx^\mu \wedge d \bar\theta) ( [ B_\mu, \Phi ] )
- (d \bar\theta \wedge d \bar\theta) (\Phi \Phi)
- (d \theta \wedge d \bar\theta)(\{\Phi, \bar \Phi \}).
\end{array} \eqno(3.3)
$$
It is interesting to point out that the condition
($ B (x) + \bar B(x) = i (C \times \bar C)$) (see, e.g., [38]), 
required for the definition of 
the anti-BRST symmetry transformations $s_{ab}$ (cf. section 2),
emerges here automatically in the superfield formulation.
In the above computation, the basic super derivative $\tilde d$ and
1-form super connection $\tilde A$ are defined on the six 
$(4 + 2)$-dimensional supermanifold, in terms of the
superspace differentials $dx^\mu, d\theta$ and  $d \bar\theta$, as
$$
\begin{array}{lcl}
\tilde d &=& d Z^M \;\partial_{M} \equiv
d x^\mu \;\partial_\mu  + d \theta\; \partial_{\theta}
+ d \bar \theta\;\partial_{\bar\theta},
\nonumber\\
\tilde  A &=& d Z^M\; (\tilde A_{M} \cdot T) \equiv
dx^\mu \;(B_\mu \cdot T) 
+ d \theta\; (\bar \Phi \cdot T)
+ d \bar \theta\; (\Phi \cdot T),
\end{array} \eqno(3.4)
$$
where $Z^M = (x^\mu, \theta, \bar\theta)$ and $\partial_M =
(\partial/\partial Z^M)$ are the generic superspace 
coordinates and the corresponding superspace derivatives on the six
$(4 + 2)$-dimensional supermanifold.
Ultimately, the insertions of the values from (3.2) and the 
use of the (anti-)BRST
transformations of equations (2.6) and (2.2) lead to the following form 
for the expansion (3.1):
$$
\begin{array}{lcl}
B_{\mu} (x, \theta, \bar \theta) &=& A_{\mu} (x) 
+ \; \theta\;  (s_{ab} A_\mu (x)) + \;\bar \theta\; (s_{b} A_\mu (x)) 
+ \;\theta\; \bar\theta \; (s_{b} s_{ab} A_\mu (x)), 
\nonumber\\
\Phi (x, \theta, \bar \theta) &=& C (x) 
+ \; \theta \; (s_{ab} C (x)) + \; \bar \theta\; (s_{b} C (x))
+ \; \theta\; \bar \theta \;(s_{b} s_{ab}  C (x)), \nonumber\\
\bar \Phi (x, \theta, \bar \theta) &=& \bar C (x) 
+ \;\theta\; (s_{ab} \bar C (x)) 
+ \; \bar \theta \; (s_{b} \bar C (x)) 
+ \;\theta \;\bar \theta \;( s_{b} s_{ab} \bar C (x)).
\end{array} \eqno(3.5)
$$
In the above equation, the shorter (group valued)
 notations (i.e. $B_\mu = B_\mu \cdot T,
\Phi = \Phi \cdot T, \bar \Phi = \bar \Phi \cdot T$ etc.) have been
used for expansions along the Grassmannian directions. It is 
clear from  (3.5) that the (anti-)BRST charges $Q_{(a)b}$ do correspond
to the translation generators $(\mbox{Lim}_{\bar\theta \rightarrow 0}
(\partial/\partial \theta)) \;\mbox {Lim}_{\theta \rightarrow 0}
(\partial/\partial\bar\theta)$ along the Grassmannian directions
of the supermanifold. 

To dwell a bit on the derivation of the on-shell nilpotent BRST symmetry
transformations (2.3) for the gauge- and the
(anti-)ghost fields, we recapitulate the bare essentials of our earlier
works [32,33]. To this end in mind, we begin with the chiral
(i.e. $\theta \rightarrow 0$) limit of (i) the expansions in (3.1), and
(ii) the definitions in (3.4) as follows [32,33] 
$$
\begin{array}{lcl}
B_{\mu}^{(c)} (x,  \bar \theta) &=& A_{\mu} (x) 
+\; \bar \theta\; R_{\mu} (x),  \qquad
\Phi^{(c)} (x,   \bar \theta) = C (x) 
+ i\; \bar \theta \;{\cal B}  (x), \nonumber\\
\bar \Phi^{(c)} (x,  \bar \theta) &=& \bar C (x) 
+ i\; \bar \theta \;B (x), \qquad
\tilde d|_{(c)} = dx^\mu \;\partial_\mu 
+ d \bar\theta\; \partial_{\bar\theta}, \nonumber\\
\tilde A|_{(c)} (x, \bar\theta) &=& d x^\mu\; B_\mu^{(c)} (x,\bar\theta)
+ d \bar\theta \; \Phi^{(c)} (x,\bar\theta),
\end{array} \eqno(3.6)
$$
where, for the sake of brevity,
 we have taken the help of the shorter version of our notations
where $B_\mu^{(c)} = (B_\mu^{(c)} \cdot T), \Phi^{(c)} =
(\Phi^{(c)} \cdot T), \bar \Phi^{(c)} = (\bar \Phi^{(c)} \cdot T)$, etc.
The application of the horizontality condition in terms of the chiral
super 1-form connection and super derivative
(i. e. $ \tilde d|_{(c)} \tilde A|_{(c)} 
+ \tilde A|_{(c)} \wedge \tilde A_{(c)}
= d A + A \wedge A$) yields the following relationships between the
secondary fields and the basic fields (see [32,33] for all the details)
$$
\begin{array}{lcl}
R_\mu (x) = D_\mu C (x),\;\;\;\; \qquad \;\;\;
{\cal B} (x) = - \frac{i}{2}\; (C \times C) (x),
\end{array} \eqno(3.7)
$$
where the individual terms on the left hand side ($ \tilde d|_{(c)} 
\tilde A|_{(c)} + \tilde A|_{(c)} \wedge \tilde A_{(c)}$)
of the above horizontality condition possess  the following explicit forms
$$
\begin{array}{lcl}
\tilde d|_{(c)} \tilde A|_{(c)} &=& 
(dx^\mu \wedge dx^\nu) (\partial_\mu B^{(c)}_\nu)
+ (dx^\mu \wedge d \bar\theta) 
(\partial_\mu \Phi^{(c)} - \partial_{\bar\theta} B^{(c)}_\mu),\nonumber\\
&-& (d\bar\theta \wedge d \bar\theta) (\partial_{\bar\theta} \Phi^{(c)}),
\nonumber\\
\tilde A|_{(c)} \wedge \tilde A|_{(c)} &=& 
(dx^\mu \wedge dx^\nu) (B^{(c)}_\mu B^{(c)}_\nu)
+ (dx^\mu \wedge d \bar\theta) ( [ B^{(c)}_\mu, \Phi^{(c)} ] ),\nonumber\\
&-& (d \bar\theta \wedge d \bar\theta) (\Phi^{(c)} \Phi^{(c)}).
\end{array} \eqno(3.8)
$$
It is evident that the above horizontality restriction does not 
shed any light on the auxiliary field $B = (B \cdot T)$. However,
the equation of motion $ B + (\partial_\mu A^\mu) = 0$, emerging
from the Lagrangian density (2.1), provides the relationship
between the auxiliary field $B$ and the basic field $A_\mu$.
The insertion of all  these values in the chiral expansion (3.6)
({\it vis-{\`a}-vis} the symmetry transformations in (2.3)) leads to
$$
\begin{array}{lcl}
B_{\mu}^{(c)} (x,  \bar \theta) &=& A_{\mu} (x) 
+\; \bar \theta\; (\tilde s_b A_\mu (x)),  \qquad
\Phi^{(c)} (x,   \bar \theta) = C (x) 
+ \bar \theta \; (\tilde s_b C (x)), \nonumber\\
\bar \Phi^{(c)} (x,  \bar \theta) &=& \bar C (x) 
+ \bar \theta \; (\tilde s_b \bar C (x)).
\end{array} \eqno(3.9)
$$
The above expansion clearly establishes that the on-shell 
($\partial_\mu D^\mu C = 0$) nilpotent $(\tilde Q_{b}^2 = 0$)
BRST charge $\tilde Q_b$ corresponds to the translation 
($ \partial/ \partial \bar\theta$) along
the $\bar\theta$-direction of the $(4 + 1)$-dimensional
chiral super sub-manifold
parametrized by the spacetime variable $x^\mu$ and a Grassmannian
variable $\bar\theta$. We would like to lay emphasis on the fact that
the anti-chiral (i.e.$\bar\theta \to 0$) limit of (i) the expansion in (3.1),
and (ii) the definitions in (3.4) does not lead to any worthwhile symmetries.
Thus, in some sense, the superfield formalism does shed some light on the
absence of the on-shell nilpotent anti-BRST symmetry
transformations  for the non-Abelian
gauge theory in any dimensions of spacetime. This observation
should be contrasted with the
Abelian gauge theory where the on-shell nilpotent (anti-)BRST symmetries
do exist for the Lagrangian density of the Abelian gauge theory
(see, e.g., [32] for details).
We shall discuss more about 
the geometrical interpretations of some of the key properties associated with
$Q_{(a)b}$ in section 5 of our present paper.\\

\noindent
{\bf 4 Nilpotent Symmetries for Dirac Fields: Augmented Superfield Approach}\\

\noindent
To obtain the nilpotent (anti-)BRST transformations for the Dirac fields
(cf. (2.6), (2.3) and (2.2)) in the framework of the augmented
superfield formalism,
we begin with the following super expansion for the superfields
$\Psi (x,\theta,\bar\theta), \bar \Psi (x,\theta,\bar\theta)$ corresponding
to the ordinary Dirac fields $ \psi (x), \bar\psi (x)$ of Lagrangian density
(2.1):
$$
\begin{array}{lcl}
\Psi (x, \theta, \bar \theta) &=& \psi (x) 
+ i\;\theta\; \; (\bar b_1 \cdot T) (x) + i \;\bar \theta\; (b_2 \cdot T) (x) 
+ i \;\theta \;\bar \theta \;(f \cdot T) (x), 
\nonumber\\
\bar \Psi (x, \theta, \bar \theta) &=& \bar \psi (x) 
+ i\;\theta\; (\bar b_2 \cdot T) (x) 
+ i \; \bar \theta \;(b_1 \cdot T) (x) 
+ i \;\theta \;\bar \theta \; (\bar f \cdot T) (x).
\end{array} \eqno (4.1)
$$
A few comments are in order now.
First, it is clear from the above that the basic tenets of supersymmetry are
satisfied here because the number of degrees of freedom of the bosonic
(i.e. $b_1, \bar b_1, b_2, \bar b_2$) and the fermionic 
(i.e. $\psi, \bar \psi, f, \bar f$)
fields do match here. Second, in general, these fields are
group valued  (i.e. $b_{1}  \equiv b_1 \cdot T, 
b_2  \equiv b_2 \cdot T, f  \equiv f \cdot T,$ etc.). Third, in the limit
$\theta, \bar\theta \rightarrow 0$, we obtain the ordinary fermionic fields
(i.e. Dirac fields) $\psi (x)$ and $\bar\psi (x)$. Fourth, unlike the
horizontality condition (i.e. $\tilde F = F$) of the previous section (which 
involves the use of (super)exterior derivatives $(\tilde d)d$ and the 
1-form (super)connections $(\tilde A)A$), we exploit here the 
invariance of the (super)currents on the (super)manifolds. To this end 
in mind, we construct here the supercurrent 
$\tilde J_\mu (x,\theta,\bar\theta)$ with the fermionic superfields of
(4.1) as
$$
\begin{array}{lcl}
\tilde J_\mu (x,\theta,\bar\theta) = \bar \Psi (x,\theta,\bar\theta)
\;\gamma_\mu\; \Psi (x,\theta,\bar\theta) 
= J_\mu (x) + \theta\; \bar K_\mu (x) + \bar\theta\; K_\mu (x)
+ i\;\theta\; \bar\theta\; L_\mu (x),
\end{array} \eqno (4.2)
$$
where the bosonic components $J_\mu (x), L_\mu (x)$ are along the
$\hat {\bf 1}$ and $\theta \bar\theta$-directions of the supermanifold
and the fermionic components $\bar K_\mu (x), K_\mu (x)$ are along the
Grassmannian directions $\theta$ and $\bar\theta$ of the supermanifold.
These components (with $J_\mu (x) = \bar\psi \gamma_\mu \psi$)
can be expressed in terms of the components of the
basic expansion (4.1) as
$$
\begin{array}{lcl}
&&\bar K_\mu (x) = i\; \bigl [(\bar b_2 \cdot T) \gamma_\mu \psi 
- \bar \psi \gamma_\mu (\bar b_1 \cdot T) \bigr ],\; \;\qquad
K_\mu (x) = i\; \bigl [(b_1 \cdot T) \gamma_\mu \psi 
- \bar \psi \gamma_\mu (b_2\cdot T) \bigr ], \nonumber\\
&& L_\mu  (x) = (\bar f\cdot T) \gamma_\mu \psi + \bar \psi 
\gamma_\mu (f \cdot T)
+ i\; \bigl [(\bar b_2 \cdot T)\gamma_\mu (b_2\cdot T) 
- (b_1 \cdot T) \gamma_\mu (\bar b_1 \cdot T) \bigr ].
\end{array} \eqno (4.3)
$$
To be consistent with the following inter-relationships
$$
\begin{array}{lcl}
s_b \leftrightarrow \mbox{Lim}_{\theta \rightarrow 0}
(\partial/\partial \bar\theta) \leftrightarrow Q_b,\; \;\;\qquad\;\;
s_{ab} \leftrightarrow \mbox{Lim}_{\bar\theta \rightarrow 0}
(\partial/\partial \theta) \leftrightarrow Q_{ab}. 
\end{array} \eqno (4.4)
$$
that were established earlier (cf. section 3) in the context of the derivation
of the off-shell  
nilpotent transformations for the 
(bosonic) gauge- and the (fermionic) (anti-)ghost fields, it
is interesting to re-express (4.2) as given below
$$
\begin{array}{lcl}
\tilde J_\mu (x,\theta,\bar\theta)
= J_\mu (x) + \theta\; (s_{ab} J_\mu (x)) + \bar\theta\; 
(s_b J_\mu (x))
+ \;\theta\; \bar\theta\; (s_b s_{ab} J_\mu (x)).
\end{array} \eqno (4.5)
$$
It can be explicitly checked, however, that $s_{(a)b} J_\mu (x) = 0$
if we exploit the exact form of nilpotent (anti-)BRST transformations of (2.6)
(2.3) and (2.2) for the matter fields $\psi$ and $\bar\psi$.
Thus, the {\it natural} restriction that emerges on the six
$(4 + 2)$-dimensional (super)manifolds is:
$\tilde J_\mu (x,\theta,\bar\theta) = J_\mu (x).$
In the physical language, this restriction implies that there are no 
superspace
(i.e. Grassmannian) contribution to the conserved supercurrent and it is
identically equal to the usual conserved current 
$J_\mu (x) = \bar \psi \gamma_\mu \psi$ on
the 4D ordinary manifold. This natural requirement, in turn, implies 
the following conditions: 
$$
\begin{array}{lcl}
&& (b_1 \cdot T) \;\gamma_\mu \;\psi = \bar \psi \;
\gamma_\mu \;(b_2 \cdot T), \;\;\;\qquad\;\;\;
(\bar b_2 \cdot T) \;\gamma_\mu \;\psi 
= \bar \psi\; \gamma_\mu \;(\bar b_1 \cdot T), \nonumber\\
&&(\bar f \cdot T)\; \gamma_\mu\; \psi + \bar \psi\; \gamma_\mu \;(f\cdot T)
= i\;\bigl [\; (b_1 \cdot T)\; \gamma_\mu \;
(\bar b_1 \cdot T) - (\bar b_2 \cdot T)\; \gamma_\mu\; (b_2 \cdot T)\; \bigr ].
\end{array} \eqno (4.6)
$$
The above conditions
are to be satisfied by the components of the basic expansion (4.1).
In other words, we have to find the solution to the above restrictions
so that the components of (4.1) could be expressed in terms of the
basic fields of the Lagrangian density (2.1) or (2.4). 
Such a solution of our interest is given below
$$
\begin{array}{lcl}
&&b_1 \equiv b_1 \cdot T = - \bar \psi (C \cdot T), \qquad
b_2 \equiv b_2 \cdot T = - (C \cdot T)\; \psi, \nonumber\\
&& \bar b_1 \equiv \bar b_1 \cdot T = -  (\bar C \cdot T) \psi, \qquad
\bar b_2 \equiv \bar b_2 \cdot T = - \bar \psi (\bar C \cdot T), \nonumber\\
&& f \equiv f \cdot T =
-\; i \;\bigl [ B \cdot T + \frac{1}{2}\; (C \times \bar C) \cdot T
\bigr ]\; \psi (x), \nonumber\\
&& \bar f \equiv \bar f \cdot T =
\; i \;\bar \psi (x)
\;\bigl [ B \cdot T + \frac{1}{2}\; (C \times \bar C) \cdot T
\bigr ]. 
\end{array} \eqno (4.7)
$$
It can be seen from 
$b_1 \gamma_\mu \psi = \bar\psi \gamma_\mu b_2$ that this equality could
be satisfied if and only if $b_1$ and $b_2$ were proportional to $\bar\psi$
and $\psi$, respectively. However, the bosonic and group-valued nature of
$b_1$ and $b_2$ enforces that the group-valued fermionic
ghost fields should be
brought in to make this equality work. This is why we take
$b_1 = - \bar \psi \;(C \cdot T)$ and
$b_2 = - (C \cdot T)\; \psi$. The rest of the choices in (4.7) are made
along similar lines of argument.
Insertion of the above values of the 
bosonic components $b_1, \bar b_1, b_2, \bar b_2$
and the fermionic components $f$ and $\bar f$ in the basic super 
expansion (4.1) leads to the following
$$
\begin{array}{lcl}
\Psi (x, \theta, \bar \theta) &=& \psi (x) 
+ \;\theta\;  (s_{ab} \psi(x)) + \;\bar \theta\; (s_b \psi (x)) 
+ \;\theta \;\bar \theta \; (s_b s_{ab}  \psi (x)), 
\nonumber\\
\bar \Psi (x, \theta, \bar \theta) &=& \bar \psi (x) 
+ \;\theta\; (s_{ab} \bar \psi (x)) 
+ \; \bar \theta \;(s_b \bar \psi (x)) 
+ \;\theta \;\bar \theta \; (s_b s_{ab} \bar \psi (x)).
\end{array} \eqno (4.8)
$$
This establishes the sanctity of the mappings in (4.4). 
In fact, these mappings are valid for all the fields
of the interacting 1-form non-Abelian gauge theory in 4D. It is also
clear that there is a mutual consistency 
and complementarity between the
horizontality condition and the invariance of the conserved matter
(super)currents of the theory on the 
most general $(4 + 2)$-dimensional (super)manifolds.

It is worthwhile to mention, before we wrap up this section, that
the off-shell nilpotent (anti-)BRST transformations for the matter fields can
be derived separately (and independently)
in a very simple manner. For this purpose, we first
take the chiral (i.e. $\theta \to 0$) limit 
of the expansion in (4.1) to obtain an expansion on a five
$(4 + 1)$-dimensional
chiral super sub-manifold, embedded in the most general
$(4 + 2)$-dimensional supermanifold, as
$$
\begin{array}{lcl}
\Psi^{(c)} (x, \bar \theta) &=& \psi (x) 
+ i\;\bar \theta\; \; (b_2 \cdot T) (x), \qquad
\bar \Psi^{(c)} (x, \bar \theta) = \bar \psi (x) 
+ i\;\bar \theta\; (b_1 \cdot T) (x). 
\end{array} \eqno (4.9)
$$
We can now construct the chiral super current, using the above expansion,
as follows
$$
\begin{array}{lcl}
\tilde J_\mu^{(c)} (x, \bar\theta) = \bar \Psi^{(c)} (x, \bar\theta)
\;\gamma_\mu\; \Psi^{(c)} (x, \bar\theta) 
= J_\mu (x) + i \;\bar\theta\; \bigl [ (b_1 \cdot T) \gamma_\mu \psi
- \bar \psi \gamma_\mu (b_2 \cdot T) \bigr ].
\end{array} \eqno (4.10)
$$
Exploiting the natural restriction, discussed earlier in great detail, that
the chiral supercurrent $\tilde J_\mu^{(c)} (x, \bar\theta)$
defined on the chiral super sub-manifold should be equal
(i.e. $\tilde J_\mu^{(c)} (x, \bar\theta) = J_\mu (x)$) to
the ordinary current $J_\mu (x)$, we obtain
the following expressions for $b_1$ and $b_2$ in terms of the basic
fields of the theory:
$$
\begin{array}{lcl}
b_1 \equiv b_1 \cdot T = - \bar \psi (C \cdot T),\; \qquad
b_2 \equiv b_2 \cdot T = - (C \cdot T)\; \psi. 
\end{array} \eqno (4.11)
$$
The substitution of these values in the chiral expansion (4.9) leads to 
the derivation of the BRST symmetry transformations for the matter fields as
$$
\begin{array}{lcl}
\Psi^{(c)} (x, \bar \theta) &=& \psi (x) 
+ \;\bar \theta\; \; (s_b \psi (x)), \qquad
\bar \Psi^{(c)} (x, \bar \theta) = \bar \psi (x) 
+ \;\bar \theta\; (s_b \bar\psi (x)). 
\end{array} \eqno (4.12)
$$
Exploiting (2.7), it is clear that the generator $Q_b$
of the off-shell nilpotent 
BRST symmetry transformations is the generator of translation
$(\partial/\partial \bar\theta)$ along the Grassmannian direction
$\bar\theta$ of the $(4 + 1)$-dimensional super sub-manifold.
In an analogous manner, one can derive the anti-BRST symmetries for the matter
fields by taking the anti-chiral 
(i.e. $\bar\theta \to 0$) limit of the super expansion in (4.1) as given below
$$
\begin{array}{lcl}
\Psi^{(ac)} (x, \theta) &=& \psi (x) 
+ i\;\theta\; \; (\bar b_1 \cdot T) (x), \qquad
\bar \Psi^{(ac)} (x, \bar \theta) = \bar \psi (x) 
+ i\;\theta\; (\bar b_2 \cdot T) (x). 
\end{array} \eqno (4.13)
$$
In terms of these expansion, the anti-chiral super current
$\tilde J_\mu^{(ac)} (x, \theta)$, defined on the $(4 + 1)$-dimensional
anti-chiral super sub-manifold, can be written as follows
$$
\begin{array}{lcl}
\tilde J_\mu^{(ac)} (x,\theta) = \bar \Psi^{(ac)} (x, \theta)
\;\gamma_\mu\; \Psi^{(ac)} (x, \theta) 
= J_\mu (x) + i \theta\; \bigl [ (\bar b_2 \cdot T) \gamma_\mu \psi
- \bar \psi \gamma_\mu (\bar b_1 \cdot T) \bigr ].
\end{array} \eqno (4.14)
$$
The natural restriction $\tilde J_\mu^{(ac)} (x,\theta) = J_\mu (x)$
on the anti-chiral $(4 + 1)$-dimensional
super sub-manifold implies the following expressions for $\bar b_1$ and 
$\bar b_2$ in terms of the basic fields of the 
Lagrangian density (2.1) of the theory:
$$
\begin{array}{lcl}
 \bar b_1 \equiv \bar b_1 \cdot T = -  (\bar C \cdot T) \psi,\; \qquad\;
\bar b_2 \equiv \bar b_2 \cdot T = - \bar \psi (\bar C \cdot T). 
\end{array} \eqno (4.15)
$$
The insertions of these values in the super expansion (4.13) leads to the
derivation of the anti-BRST symmetry transformations for the matter fields
as illustrated below
$$
\begin{array}{lcl}
\Psi^{(ac)} (x, \bar \theta) &=& \psi (x) 
+ \; \theta\; \; (s_{ab} \psi (x)), \qquad 
\bar \Psi^{(ac)} (x, \bar \theta) = \bar \psi (x) 
+ \; \theta\; (s_{ab} \bar \psi (x)). 
\end{array} \eqno (4.16)
$$
This establishes the geometrical interpretation for the nilpotent 
anti-BRST charge $Q_{ab}$ as the translation generator
$(\partial/\partial\theta)$ along the Grassmannian $\theta$-direction of the
anti-chiral $(4 + 1)$-dimensional
super sub-manifold. Physically, the process of translation of the
super fields $\Psi (x,\theta)$ and 
$\bar \Psi (x, \theta)$ along the $\theta$-direction of the 
five $(4 + 1)$-dimensional anti-chiral
super sub-manifold generates the off-shell nilpotent
anti-BRST symmetry transformations for the
ordinary Dirac fields $\psi (x)$ and $\bar\psi (x)$ of the
Lagrangian density (2.1) of the theory.\\

\noindent
{\bf 5 Conclusions}\\

\noindent
We have exploited in our present investigation
(i) the horizontality condition, and (ii) the invariance of the conserved 
matter (super)currents on a set of (super)manifolds. These supermanifolds
are of (1) a general nature with $(4 + 2)$-dimensional superspace variables
$Z^M = (x^\mu,\theta,\bar\theta)$, and (2) a special variety with
$(4 + 1)$-dimensional (anti-)chiral superspace variables $Z^M = (x^\mu,\theta)$
and/or $Z^M = (x^\mu,\bar\theta)$.
The above physically motivated restrictions
(i) and (ii), which are the salient features
of the  augmented superfield formulation, lead to primarily
{\it four} key consequences. First, they provide the geometrical 
interpretation for the conserved and off-shell
nilpotent (anti-)BRST charges $Q_{(a)b}$ as the translation generators
(($\mbox{Lim}_{\bar\theta \rightarrow 0} (\partial/\partial\theta),
\mbox{Lim}_{\theta \rightarrow 0} (\partial/\partial\bar\theta)$)
along the Grassmannian directions $(\theta)\bar\theta$ of the 
most {\it general} $(4 + 2)$-dimensional supermanifold.
Second, they produce {\it together} the off-shell
nilpotent (anti-)BRST symmetry transformations
for the gauge-, the (anti-)ghost-  and the matter fields of
an interacting gauge theory. Third, 
the anticommutativity ($s_b s_{ab} + s_{ab} s_b = 0$) of the (anti-)BRST
symmetries (as well as corresponding charges) is expressed
in the language of the
translation generators because $(\partial/\partial \theta)
(\partial/\partial \bar\theta) + (\partial/\partial \bar\theta)
(\partial/\partial \theta) = 0$. Finally, they furnish the geometrical
interpretation for the nilpotency property (i.e. $s_{(a)b}^2 = 0$) of the
(anti-)BRST transformations (and their corresponding nilpotent generators)
in terms of a couple of successive translations
(i.e. $(\partial/\partial\theta)^2 = 0,
(\partial/\partial\bar\theta)^2 = 0$) along either
of the Grassmannian directions $(\theta)\bar\theta$ of the 
$(4 + 2)$-dimensional supermanifold.

As a side remark,
it is interesting to point out that the {\it on-shell} nilpotent version 
of the symmetry transformations for the gauge- and 
the (anti-)ghost fields have been derived
by exploiting the horizontality condition on the $(4 + 1)$-dimensional
(anti-)chiral super sub-manifolds, 
embedded in the general $(4 + 2)$-dimensional
supermanifold. For instance, on the $(4 + 1)$-dimensional chiral super 
sub-manifold, the 
on-shell ($\partial_\mu D^\mu C = 0$) nilpotent ($\tilde Q_b^2 = 0$)
BRST charge $\tilde Q_b$ for the non-Abelian gauge theory
corresponds to the translation generator $(\partial/\partial\bar\theta)$ 
along the $\bar\theta$-direction of the super sub-manifold
(cf. section 3). There exists no such on-shell nilpotent anti-BRST 
symmetry in the present non-Abelian gauge theory (as emphasized
in section 2). The superfield formulation sheds some light on the
reasons behind the non-existence of the on-shell nilpotent anti-BRST 
symmetries for the non-Abelian gauge theory (see, e.g., [32] for details). 
In contrast, the requirement
of the invariance of the (super)currents on (i) the $(4 + 2)$-dimensional
supermanifold, and (ii) the $(4 + 1)$-dimensional (anti-)chiral super
sub-manifolds, leads to the derivation of the {\it off-shell}
 nilpotent and anticommuting (anti-)BRST
symmetry transformations for the matter fields of the interacting non-Abelian
gauge theory. In the first case, we obtain these (anti-)BRST
symmetries together and, in the latter case, we obtain
these symmetries separately and independently. The geometrical interpretation
for the (anti-)BRST charges $Q_{(a)b}$ as the translation generators along
$(\theta)\bar\theta$-directions of the supermanifold remains intact for
the cases of the general supermanifold as well as  
the (anti-)chiral supermanifolds.

One of the most interesting features of the above restrictions on the
supermanifolds is the mutual consistency and complementarity between them.
These restrictions are at the heart of a  complete geometrical description of
all the nilpotent symmetry transformations for all the fields present in an 
interacting 1-form non-Abelian gauge theory. It is worthwhile to mention
that, physically, the invariance 
($\tilde J_\mu (x,\theta,\bar\theta) = J_\mu (x)$) of the gauge invariant
and conserved matter (super)currents on the (super)manifolds
implies the conservation of charge which, ultimately,
does not get any contribution from the superspace (Grassmannian) variables.
This restriction turns out to be the {\it natural} one on the supermanifold. 
On the other hand,
physically, the horizontality condition owes its origin to the
gauge invariance of the electric and magnetic fields for the Abelian
gauge theory [28,29] and the gauge invariance of the kinetic energy
term $(- \frac{1}{4} F^{\mu\nu} \cdot F_{\mu\nu})$ for the case of the
non-Abelian gauge theory. Thus, the horizontality condition ($\tilde F = F$)
requires that the Grassmannian (i.e. $\theta$ and $\bar\theta$)
contribution to the 2-form curvature field
should be {\it zero} because the second-rank
tensor associated with it corresponds to the gauge invariant physical
electric- and magnetic fields. Mathematically, the equality $\tilde F = F$
owes its origin to the
cohomological (super) exterior derivatives $(\tilde d)d$
which play a very decisive role in the definition of the (super) 2-forms
$(\tilde F)F$. In this connection, it is pertinent to 
point out that, in
a recent set of papers [39,40], all the three 
\footnote{On an ordinary manifold without a boundary, the operators
$(d, \delta, \Delta)$ form a set which is popularly known as the
de Rham cohomological operators of differential geometry. The
operators $(\delta)d$ are called as the (co-)exterior derivatives
(with $ d = dx^\mu \partial_\mu, \delta = \pm * d *, d^2 = \delta^2 = 0,
* =$ Hodge duality operation on the manifold) 
and $\Delta = (d + \delta)^2 = d \delta
+ \delta d$ is known as the Laplacian operator (with $ [\Delta, d] =
[\Delta, \delta] = 0$) [11-15].}
super cohomological
operators $\tilde d, \tilde \delta, \tilde \Delta$ have been exploited
in the generalized versions of the horizontality condition for the derivation
of the nilpotent (anti-)BRST, (anti-)co-BRST and a bosonic symmetry
\footnote{ This symmetry $s_w$ (with $s_w^2 \neq 0$)
is equal to the anticommutator (i.e. $s_w = \{ s_b, s_d \}
= \{ s_{ab}, s_{ad} \}$) of the nilpotent ($s_{(a)b}^2 = s_{(a)d}^2
= 0$) (anti-)BRST $s_{(a)b}$ and (anti-)co-BRST 
$s_{(a)d}$ transformations [39,40].}
transformations for the two-dimensional (2D) free Abelian gauge theory
defined on the four $(2 + 2)$-dimensional supermanifold.

In a very
recent paper [41], the analogue of the Hodge duality $*$ operation has
been defined on the six $(4 + 2)$-dimensional supermanifold
by exploiting the mathematical power of $\tilde d, \tilde \delta$.
Furthermore, the
non-local, non-covariant and nilpotent dual(co)-BRST symmetry transformations
for the gauge- and the (anti-)ghost fields
have been obtained in the framework of six $(4 + 2)$-dimensional superfield
formulation for the 4D Abelian gauge theory [41]. However, the non-local,
non-covariant and nilpotent transformations for the {\it matter} (Dirac)
fields, that exist in literature (see, e.g., [42,43] for details),
have not yet been derived for the interacting Abelian gauge theory
in superfield formulation (see, e.g., [41] for details). It would be a very 
nice endeavour to obtain these nilpotent symmetries for the matter fields
by exploiting the invariance of the {\it non-local}
(super)currents on the six-dimensional (super)manifolds
in the framework of the augmented superfield formalism. Such kind
of nilpotent symmetries for the matter fields
also exist for the interacting 4D non-Abelian gauge theory [43].
It would be a very challenging endeavour to capture these nilpotent symmetries
in the framework of the six $(4 + 2)$-dimensional superfield formalism
for the 4D interacting non-Abelian gauge theory in the general scheme of the
augmented superfield formulation. These are some of the issues that 
are under investigation at the moment.

\end{document}